# A new approach to single-tone frequency estimation via linear least squares curve fitting

Solomon Davis[a], Izhak Bucher[a]

[a]Technion – Israel Institute of Technology, Haifa, Israel, 3200003

**Abstract**

Presented is a new algorithm for estimating the frequency of a single-tone noisy signal using linear least squares (LLS). Frequency estimation is a nonlinear problem, and typically, methods such as Nonlinear Least Squares (NLS) (batch) or a digital phase locked loop (DPLL) (online) are employed for such an estimate. However, with the linearization approach presented here, one can harness the efficiency of LLS to obtain very good estimates, while experiencing little penalty for linearizing. In this paper, the mathematical basis of this algorithm is described, and the bias and variance are analyzed analytically and numerically. With the batch version of this algorithm, it will be demonstrated that the estimator is just as good as NLS. But because LLS is non recursive, the estimate it produces much more efficiently than from NLS. When the proposed algorithm is implemented online, it will be demonstrated that performance is comparable to a digital phase locked loop, with some stability and tracking range advantages.

**Section 1: Introduction**

The presented algorithm allows one to estimate the frequency of a noisy, single tone signal with an excellent combination of accuracy and speed through a novel application of Linear Least Squares (LLS) [1]. Such an estimator has numerous uses, such as measuring the Doppler shift from ultrasonic [2] or radar signals [3], measuring the natural frequency shift of a microcantilever in atomic force microscopy [4] and others. Many techniques already exist for estimating the frequency of a noisy single tone signal. Some popular batch techniques are the Fast Fourier Transform [5] MUSIC [6], and Nonlinear Least Squares (NLS) [7]. For online estimation there are also a number of options such as Maximum Likelihood [8], Autocorrelation [9], the Hilbert Transform [10], or a digital phase locked loop (DPLL) [11].

The proposed LLS-based method can be implemented in either batch method or online versions, and appears to experience some advantages over the abovementioned techniques. For example, when the batch version is implemented, it will be shown that this algorithm produces almost identical results to NLS, but performs the calculation much more efficiently because it is non recursive. Furthermore, when the online version is implemented, it will be shown that performance is comparable to that of a DPLL, but is not dependent on a Lock-in time, Lock-in-range or Hold-in range [11]. This versatility is a significant advantage when the frequency of a single tone signal need be tracked over a wide frequency range, but where accuracy and efficient data utilization are also essential.

The first part of this paper develops analytically and numerically the method of estimating the frequency of a noisy single tone signal using LLS. The performance of the algorithm (the variance and bias of the estimator) in the presence of white noise will be derived. The second half of this paper uses simulation to validate the predicted performance. The performance will also be compared with that of NLS and a DPLL.

**Section 2: Algorithm Structure**

The structure of the algorithm consists of two "stages" of LLS. The first stage makes a number of instantaneous phase estimates of the noisy signal. If the frequency of this signal is constant, then the phase will progress linearly. Once a number of phase estimates are made, the second stage uses LLS again to fit their time dependent slope, which is an estimate of the signal's frequency.

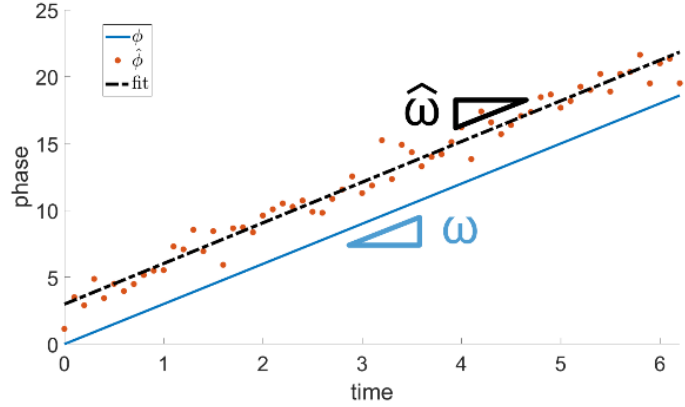

Figure 1. Estimating $\hat{\omega}$ by fitting a slope to the instantaneous phase estimate vector $\hat{\boldsymbol{\phi}}$.

The novel step of this algorithm is the use of LLS to estimate the instantaneous phase of the signal. But like frequency, this problem is also nonlinear, and again it seems LLS cannot be used. However, it will be demonstrated that if the signal frequency is known in advance within roughly 10 percent, it is possible to obtain good instantaneous phase estimates with very little penalty.

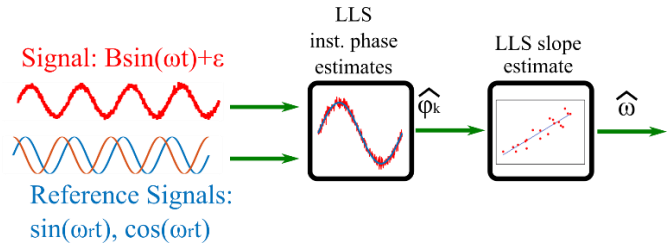

Figure 2. Diagram of the frequency estimation algorithm with the two LLS stages.

**Section 3: The Single Tone Signal with Additive White Gaussian Noise**

Our observed signal is of the form [1]:

$$y_i = \beta_1 \sin(\omega t_i) + \beta_2 \cos(\omega t_i) + \varepsilon_i = B\sin(\omega t_i + \theta) + \varepsilon_i \qquad (1)$$

The signal in matrix form is

$$\boldsymbol{y} = \begin{bmatrix} y_1 \\ \vdots \\ y_n \end{bmatrix}, \quad \mathbf{A} = \begin{bmatrix} \sin(\omega t_i) & \cos(\omega t_i) \\ \vdots & \vdots \\ \sin(\omega t_n) & \cos(\omega t_n) \end{bmatrix}, \quad \boldsymbol{\beta} = \begin{bmatrix} \beta_1 \\ \beta_2 \end{bmatrix}, \quad \boldsymbol{\varepsilon} = \begin{bmatrix} \varepsilon_1 \\ \vdots \\ \varepsilon_n \end{bmatrix}, \qquad (2)$$

and written compactly as

$$y = \mathbf{A}\boldsymbol{\beta} + \boldsymbol{\varepsilon} .\tag{3}$$

It is assumed the measurement noise $\boldsymbol{\varepsilon}$ follows the multivariate normal distribution:

$$\boldsymbol{\varepsilon} \sim \mathsf{N}(\boldsymbol{\mu}, \boldsymbol{\Sigma}) = \mathsf{N}(\mathbf{0}, \sigma_\varepsilon^2 \mathbf{I})\tag{4}$$

where

$$\begin{aligned}\boldsymbol{\mu} &= \begin{bmatrix} E\{\varepsilon_1\} & \cdots & E\{\varepsilon_n\} \end{bmatrix}^T = \begin{bmatrix} \mu_1 & \cdots & \mu_n \end{bmatrix}^T \\ \boldsymbol{\Sigma} &= \begin{bmatrix} Cov\{\varepsilon_i, \varepsilon_j\}; 1 \le i, j \le k \end{bmatrix}\end{aligned}.\tag{5}$$

The instantaneous phase of the signal at $t_0$ is

$$\phi = \tan^{-1}\left(\frac{\beta_2}{\beta_1}\right).\tag{6}$$

**Section 4: Stage 1, Instantaneous Phase Estimator $\hat{\phi}$**

If $\mathbf{b}_\phi = \begin{bmatrix} b_{\phi,1} & b_{\phi,2} \end{bmatrix}^T$ is the estimated model of $\boldsymbol{\beta}$, the estimated model can be expressed as:

$$\mathbf{y} = \mathbf{A}_\phi \mathbf{b}_\phi + \mathbf{e}\tag{7}$$

where $\mathbf{e}$ denotes the residuals.

Now attempt to generate $\mathbf{b}_\phi$ using LLS. But the frequency $\omega$ is unknown, and for guessed frequency $\omega_r$, the model to be fitted is

$$\mathbf{A}_\phi = \begin{bmatrix} \sin(\omega_r t_1) & \cos(\omega_r t_1) \\ \vdots & \vdots \\ \sin(\omega_r t_n) & \cos(\omega_r t_n) \end{bmatrix}.\tag{8}$$

By [1], the estimated weights are

$$\mathbf{b}_\phi = \left(\mathbf{A}_\phi^{\mathsf{T}} \mathbf{A}_\phi\right)^{-1} \mathbf{A}_\phi^{\mathsf{T}} \mathbf{y} .\tag{9}$$

By (6) the estimate of the instantaneous phase at $t_0$ is

$$\hat{\phi} = \tan^{-1}\left(\frac{b_{\phi,2}}{b_{\phi,1}}\right).\tag{10}$$

By doing this many times, ones can obtain a vector of unwrapped phase estimates $\hat{\boldsymbol{\phi}} = \begin{bmatrix} \hat{\phi}_1 & \hat{\phi}_2 & \cdots & \hat{\phi}_N \end{bmatrix}^T$, which will be used in the second stage to fit their slope.

This method is called sinusoidal amplitude estimation [12]. Typically however, $\omega_r = \omega$. But it will be shown in later sections that even when $\omega_r \approx \omega \pm 10\%$, there may be very little penalty in estimating $\omega$ when $y$ is fitted with the wrong frequency.

**Section 5: Approximation of** $Var\{\hat{\phi}\}$

To develop the covariance matrix of $\mathbf{b}_\phi$,

$$\Sigma_\phi = \left(\mathbf{A}_\phi^T \mathbf{A}_\phi\right)^{-1} \sigma_\varepsilon^2, \tag{11}$$

first calculate the following

$$\mathbf{A}_\phi^T \mathbf{A}_\phi = \begin{bmatrix} \sum_{i=1}^n \sin^2(\omega_r t_i) & \sum_{i=1}^n \sin(\omega_r t_i)\cos(\omega_r t_i) \\ \sum_{i=1}^n \sin(\omega_r t_i)\cos(\omega_r t_i) & \sum_{i=1}^n \cos^2(\omega_r t_i) \end{bmatrix}, \tag{12}$$

where $n$ is the number of samples used per phase estimate.

To obtain a closed form expression, one can approximate the summations with integrals [13]:

$$\sum_{i=1}^n f(t_i) \approx F_s \int_0^{t_n} f(t)dt, \tag{13}$$

for measurement sample rate $F_s = \dfrac{1}{t_{i+1} - t_i}$, and time per phase estimate $t_n$. This provides the following approximation of (12):

$$\mathbf{A}_\phi^T \mathbf{A}_\phi \approx \begin{bmatrix} F_s \int_0^{t_n} \sin^2(\omega_r t)dt & F_s \int_0^{t_n} \sin(\omega_r t)\cos(\omega_r t)dt \\ F_s \int_0^{t_n} \sin(\omega_r t)\cos(\omega_r t)dt & F_s \int_0^{t_n} \cos^2(\omega_r t)dt \end{bmatrix}$$

$$= F_s \begin{bmatrix} \dfrac{t_n}{2} - \dfrac{\sin(2\omega_r t_n)}{4\omega_r} & \dfrac{\cos(2\omega_r t_n)}{4\omega_r} - \dfrac{1}{4\omega_r} \\ \dfrac{\cos(2\omega_r t_n)}{4\omega_r} - \dfrac{1}{4\omega_r} & \dfrac{t_n}{2} + \dfrac{\sin(2\omega_r t_n)}{4\omega_r} \end{bmatrix} \approx F_s \begin{bmatrix} \dfrac{t_n}{2} & 0 \\ 0 & \dfrac{t_n}{2} \end{bmatrix}. \tag{14}$$

Approximation (14) is valid when the estimation time $t_n$ covers many cycles of the reference waveform. i.e. $t_n \gg 2\pi / \omega_r$. This is often the case, especially when estimating high frequencies.

Taking the inverse of (14) one obtains the covariance matrix of $\mathbf{b}_\phi$

$$\Sigma_\phi \approx \dfrac{2\sigma_\varepsilon^2}{F_s t_n} \begin{bmatrix} 1 & 0 \\ 0 & 1 \end{bmatrix}. \tag{15}$$

To calculate the variance of $\hat{\phi}$ (10), approximate using the Taylor series expansion [14]

$$Var\{\hat{\phi}(b_1,b_2)\} = Var\left\{\tan^{-1}\left(\frac{b_2}{b_1}\right)\right\} \approx \sum_{p=1}^{2}\left(\frac{d\hat{\phi}(\mu_{b1},\mu_{b2})}{db_p}\right)^2 \sigma_i^2 + \sum_{p=1}^{2}\sum_{q=1,q\neq p}^{2}\left(\frac{d\hat{\phi}(\mu_{b1},\mu_{b2})}{db_p}\right)\left(\frac{d\hat{\phi}(\mu_{b1},\mu_{b2})}{db_q}\right)Cov\{b_p,b_q\} \quad (16)$$

Noting from (15) that $Cov\{b_1,b_2\} = 0$, $\sigma_i^2 = \Sigma_{ii} = \dfrac{2\sigma_\varepsilon^2}{F_s t_n}$, for mean values $\mu_{b1} = B\sin(\phi)$, $\mu_{b2} = B\cos(\phi)$, $\hat{\phi}(\mu_{b1},\mu_{b2}) = \tan^{-1}(\mu_{b2}/\mu_{b1})$,

$$\sigma_{\hat{\phi}}^2 \triangleq Var\{\hat{\phi}\} \approx \frac{2\sigma_\varepsilon^2}{B^2 F_s t_n} = \frac{1}{F_s t_n SNR} \quad (17)$$

where $\sigma_y^2 = B^2/2$ is the variance of the single tone signal without noise, and the signal to noise ratio is $SNR = \sigma_y^2/\sigma_\varepsilon^2$.

### Section 6: Stage 2, The Frequency estimate $\hat{\omega}$

Once a number of phase estimates $\hat{\phi}$ have been collected, they can be used to obtain a frequency estimate $\hat{\omega}$. These phase estimates are of the assumed form

$$\hat{\boldsymbol{\phi}} = \omega \mathbf{t} + \phi_0 + \boldsymbol{\varepsilon}_{\hat{\phi}}, \quad (18)$$

where $\boldsymbol{\varepsilon}_{\hat{\phi}}$ is the "noise" of the phase estimates with covariance matrix $\sigma_{\hat{\phi}}^2 \mathbf{I}$. To estimate $\hat{\omega}$, LLS is employed a second time with the affine model

$$\mathbf{A}_\omega = [\mathbf{1} \quad \mathbf{t}], \mathbf{1}, \mathbf{t} \in \mathbb{R}^{N \times 1}, \quad (19)$$

where $\mathbf{t} = [t_1 \quad \ldots \quad t_N]^T$ is the time vector. The fitted offset and slope are

$$\mathbf{b}_\omega = \left(\mathbf{A}_\omega^T \mathbf{A}_\omega\right)^{-1} \mathbf{A}_\omega^T \hat{\boldsymbol{\phi}}. \quad (20)$$

Therefore, the frequency estimator is the slope

$$\hat{\omega} = b_{\omega,2}. \quad (21)$$

### Section 7: Approximation of $Var\{\hat{\omega}\}$

To approximate $Var\{\hat{\omega}\}$, calculate the covariance matrix $\Sigma_\omega \in \mathbb{R}^{2\times 2}$ [1].

$$\Sigma_\omega = \left(\mathbf{A}_\omega^T \mathbf{A}_\omega\right)^{-1} \sigma_{\hat{\phi}}^2. \quad (22)$$

To find $\Sigma_\omega$ first calculate the following:

$$\mathbf{A}_\omega^T \mathbf{A}_\omega = \begin{bmatrix} N & \sum_{k=1}^{N} t_k \\ \sum_{k=1}^{N} t_k & \sum_{k=1}^{N} t_k^2 \end{bmatrix}, \quad (23)$$

where $N$ is the number of phase estimates used per frequency estimate. Using approximation (13), one obtains

$$\mathbf{A}^T \mathbf{A} \approx \begin{bmatrix} N & F_\phi \int_0^{t_N} t\, dt \\ F_\phi \int_0^{t_N} t\, dt & F_\phi \int_0^{t_N} t^2\, dt \end{bmatrix} = \begin{bmatrix} N & F_\phi \dfrac{t_N^2}{2} \\ F_\phi \dfrac{t_N^2}{2} & F_\phi \dfrac{t_N^3}{3} \end{bmatrix}, \quad (24)$$

where $t_N$ is the time per frequency estimate, and

$$F_\phi = \frac{1}{t_n} \quad (25)$$

is the rate that individual $\hat{\phi}_k$ are acquired. Taking the inverse of matrix (24), the covariance matrix is:

$$\Sigma_\omega \approx \frac{12\sigma_{\hat{\phi}}^2}{F_\phi t_N^3} \begin{bmatrix} \dfrac{t_N^2}{3} & -\dfrac{t_N}{2} \\ -\dfrac{t_N}{2} & 1 \end{bmatrix}, \quad (26)$$

Therefore, one obtains the approximation of the variance of the slope:

$$Var(\hat{\omega}) \approx \Sigma_{\omega,22} = \frac{12\sigma_{\hat{\phi}}^2}{F_\phi t_N^3}. \quad (27)$$

This is the general form of the variance of the slope estimate using LLS, with the affine model (19), and for measurement noise variance $\sigma_{\hat{\phi}}^2$.

Inserting (17) and (25) into (27),

$$Var(\hat{\omega}) \approx \frac{12}{F_s t_N^3 SNR}. \quad (28)$$

**Section 8: Effective Linearization to minimize $Bias(\hat{\omega})$**

Any bias of $\hat{\phi}$ and hence $\hat{\omega}$ is due to the difference between $\omega$ and $\omega_r$. Obviously, we wish to minimize this bias. In this section it will be shown that even if there is considerable distance between these two frequencies, the bias can be made very small or even negligible.

The expression for the bias of the estimator $\hat{\phi}$ is [15]

$$\text{Bias}\{\hat{\phi}\} = E\{\hat{\phi}\} - \phi \tag{29}$$

where $E\{\bullet\}$ is the expected value on $\bullet$.

Now approximate the expected value of the phase estimator so that the bias can be analyzed [14].

$$E\{\hat{\phi}(\varepsilon_1, \varepsilon_2 ...)\} \approx \hat{\phi}(\mu_1, \mu_2 ...) = \hat{\phi}(\mathbf{0}) = \tan^{-1}\left(\frac{b_2(\mathbf{0})}{b_1(\mathbf{0})}\right) \tag{30}$$

A closed form expression of $E\{\hat{\phi}\}$ can be seen in Appendix 1. From this expression it's observed that the estimator is of the general form

$$E\{\hat{\phi}\} = \tan^{-1}\left(\frac{A\cos(\phi) + B\sin(\phi)}{C\cos(\phi) + D\sin(\phi)}\right). \tag{31}$$

An unbiased estimate of $\hat{\omega}$ will be obtained if $\frac{d}{dt}E\{\hat{\phi}\} = \frac{d}{dt}\phi = \omega$, where any offset $E\{\hat{\phi}_0\}$ is irrelevant. Therefore, an unbiased estimate of $\omega$ will be obtained if

$$\begin{aligned}|A| &= |D| \\ |B| &= |C|\end{aligned}, \tag{32}$$

i.e. the numerator and the denominator are equal in amplitude but shifted in phase by $\pi/2$ with respect to $\phi$.

In Section 5, $t_n$ was defined as the time per phase estimate. But it will be seen that selection of $t_n$ greatly affects $\text{Bias}(\hat{\phi})$. Now define a new variable, $q$, which is a fraction of a cycle of the reference waveform used to estimate a $\hat{\phi}_k$.

$$\frac{2\pi q}{\omega_r} = t_n$$

At this point we wish to select a value of $q$ where condition (32) will be satisfied as best as possible. First selecting $q = .5$ and $q = 1$, and then inserting into Appendix 1, one obtains

$$E\{\hat{\phi}\}\Big|_{q=.5} = \tan^{-1}\left(-\left(\frac{\omega}{\omega_r}\right)\frac{\cos[\phi] + \cos\left[\phi + \frac{\pi\omega}{\omega r}\right]}{\sin[\phi] + \sin\left[\phi + \frac{\pi\omega}{\omega r}\right]}\right).$$

$$E\{\hat{\phi}\}\Big|_{q=1} = \tan^{-1}\left(\left(\frac{\omega}{\omega_r}\right)\tan\left[\phi + \frac{\pi\omega}{\omega r}\right]\right) \tag{33}$$

From expressions (33) one can see how the estimation problem has now been "linearized". If the ratio $\omega/\omega_r = 1$, then the expression within the arctan function is exactly of the from (31), (32), and $\frac{d}{dt}E\{\hat{\phi}\} = \frac{d}{dt}\phi$.

Furthermore, for $\omega/\omega_r$ near unity, (31), (32) are still approximately true, and $\frac{d}{dt}E\{\hat{\phi}\} \approx \frac{d}{dt}\phi$. However, for $\omega/\omega_r$ far from unity, $dE\{\hat{\phi}\}/dt$ will be warped and a non-negligible $Bias(\hat{\omega})$ will be present. But there is a value of $q$ that produces an even more efficient linearization.

Instead, inserting $q^* = 1/\sqrt{2}$, the expression for $E\{\hat{\phi}\}\big|_{q=1/\sqrt{2}}$ can be seen in Appendix 2. Unlike expressions (33), Appendix 2 does not collapse into a concise expression. The reason $q^*$ was chosen is because it has been observed numerically to minimize any warping in $dE\{\hat{\phi}\}/dt$ much better than $q = .5, 1$ when $\omega$ and $\omega_r$ are far apart. This was observed for a wide range of scenarios tested numerically. Such slope warping can be seen in Fig.3.

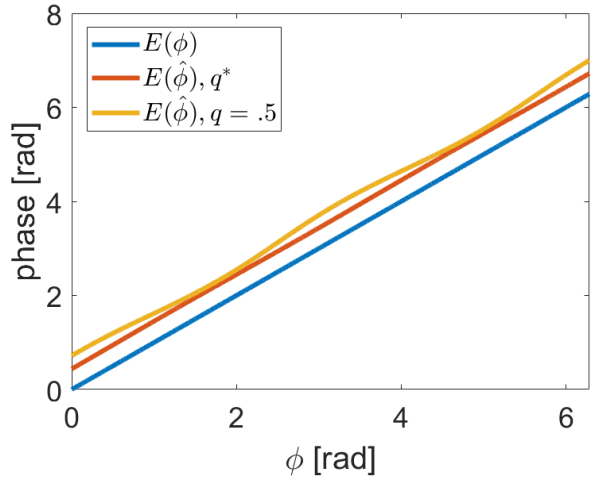

Figure 3. $E\{\hat{\phi}\}$ vs. $\phi$, $q = .5, 1/\sqrt{2}$, $f = 20kHz$, $f_r = 16kHz$, $\varepsilon_i = 0$. It's observed that there is very little warping of $dE\{\hat{\phi}\}/dt$ when $q^*$ is used.

In Fig.4 one can see the periodic bias of $\hat{\phi}$ when different values of $q$ are used.

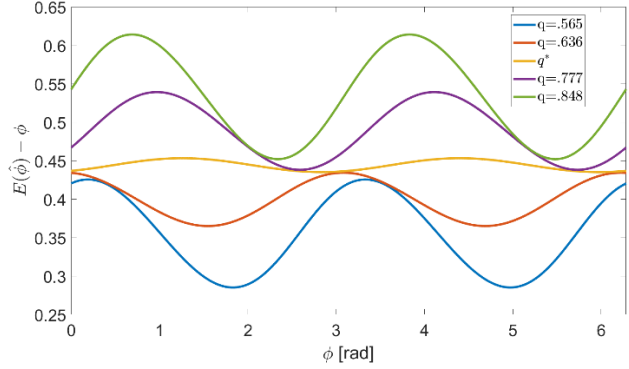

Figure 4. $Bias\{\hat{\phi}\} = E\{\hat{\phi}\} - \phi$ for various values of $q$, $f = 18000 Hz$, $f_r = 181800$

From Figs.3,4 it can be observed that $Bias\{\hat{\phi}\}$ consists of two components: 1) A constant component which will have no effect on the slope $dE\{\hat{\phi}\}/dt$ or $\hat{\omega}$. 2) A component that influences the slope of periodically with respect to $\phi$. This will add to any error of $\hat{\omega}$. Therefore, to obtain the best $\hat{\omega}$, we wish to minimize the periodic component of $Bias\{\hat{\phi}\}$, i.e. we wish to minimize $Var\{Bias\{\hat{\phi}\}\}$. From Fig.4 it can be observed that variance of the bias is minimized near $q^*$. One can visualize $Var\{Bias\{\hat{\phi}\}\}$ for various values of $q$ in Fig.5.

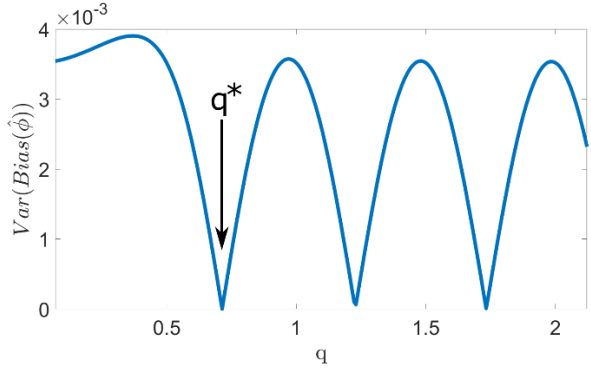

Figure 5. $Var\{Bias\{\hat{\phi}\}\}$ vs. $q$, $f_r = 300000\ Hz$, $f = 300300\ Hz$.

It's interesting to note in Fig.5, that $Bias(\hat{\phi})$ is minimized for many values of $q$, and $q^* \approx 1/\sqrt{2} + N/2$, for integers $N$.

### Section 9: Efficient Online Implementation

Both stages use LLS for their estimations, though calculating the estimators using the "batch" approach is not practical. Using Recursive Linear Least Squares (RLLS) [16] is a better option, but an even more efficient method to realize LLS is presented here. Like RLLS, this method performs the LLS as the data streams in.

It will first be shown how to realize the first LLS stage $\hat{\phi}$. Consider the matrix $\mathbf{J}_\phi$, which is calculated beforehand.

$$\mathbf{J}_\phi = \left(\mathbf{A}_\phi^T \mathbf{A}_\phi\right)^{-1} = \begin{bmatrix} J_{\phi,11} & J_{\phi,12} \\ J_{\phi,21} & J_{\phi,22} \end{bmatrix} \quad (34)$$

It can be shown that

$$b_{\phi,1} = J_{\phi,11} \sum_{i=1}^{n} \sin(\omega_r t_i) y_i + J_{\phi,1,2} \sum_{i=1}^{n} \cos(\omega_r t_i) y_i$$
$$b_{\phi,2} = J_{\phi,21} \sum_{i=1}^{n} \sin(\omega_r t_i) y_i + J_{\phi,22} \sum_{i=1}^{n} \cos(\omega_r t_i) y_i \quad (35)$$

From (35), the coefficients $b_{\phi,1}$ and $b_{\phi,2}$ can be realized efficiently by generating $\sin(\omega_r t_i)$ and $\cos(\omega_r t_i)$. A block diagram of this realization can be seen in Fig.6. Furthermore, by relation (10) one can use these coefficients to arrive at $\hat{\phi}_k$. However, the arctan function is moved to the next stage because it need not be calculated at the rate $F_s$.

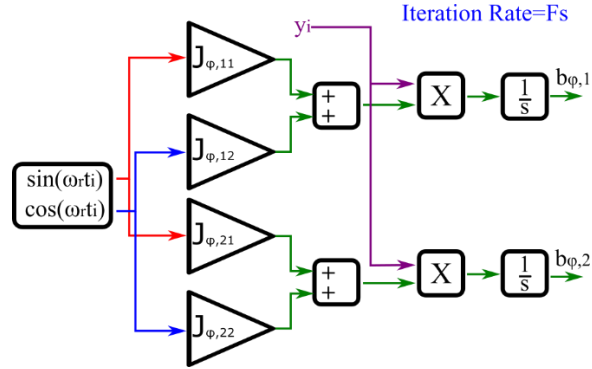

Figure 6. Block diagram of efficient code of the first stage for generating $b_{\phi,1}$ and $b_{\phi,1}$, iteration rate $F_s$ for $t_n$ seconds per estimate.

Now it will be shown how to realize the second LLS stage which calculates $\hat{\omega}$ efficiently by fitting an affine model to $\hat{\phi}$. Consider the matrix $\mathbf{J}_\omega$, which can also be calculated beforehand.

$$\mathbf{J}_\omega = \left(\mathbf{A}_\omega^T \mathbf{A}_\omega\right)^{-1} = \begin{bmatrix} J_{\omega,11} & J_{\omega,12} \\ J_{\omega,21} & J_{\omega,22} \end{bmatrix} \quad (36)$$

It can be shown that

$$\hat{\omega} = \sum_{k=1}^{N} \left(J_{\omega,22} t_k + J_{\omega,21}\right) \hat{\phi}_k \quad . \quad (37)$$

A block diagram for an efficient realization of $\hat{\omega}$ can be seen in Fig.7. In this diagram it's observed that the arctan function has been moved to the second stage.

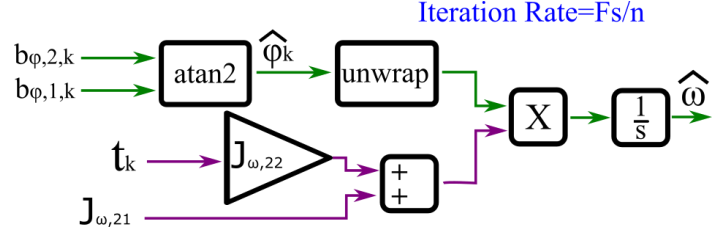

Figure 7. Block diagram of efficient code of the second stage for generating $\hat{\omega}$.

## Section 10: Simulations of $Bias\{\hat{\omega}\}$

The $Bias\{\hat{\omega}\}$ was tested online in simulation using the implementation described in the previous section. In these tests, the input signal $y$ did not have any additive noise, and therefore, the inaccuracies of $\hat{\omega}$ were only due to the difference between $\omega$ and $\omega_r$. As predicted in Section 6, $Bias\{\hat{\omega}\}$ should be minimized with $q^*$. This was the value chosen for these simulations. The results of the simulations over different parameters can be seen below.

| $f$ [Hz] | $f_r$ [Hz] | $\Delta f = f - f_r$ [Hz] | $t_N$ [$\mu s$] | $Mean\{Bias\{\hat{f}\}\}$ [Hz] | $Var\{Bias\{\hat{f}\}\}$ [$Hz^2$] |
|---|---|---|---|---|---|
| 32320 | 32e3 | 320 | 500 | -.023 | .00011 |
| 31680 | 32e3 | -320 | 500 | 0 | .00019 |
| 323200 | 320e3 | 3200 | 500 | -.001 | 7.05e-5 |
| 352000 | 320e3 | 32000 | 500 | -.001 | .00276 |
| 288000 | 320e3 | -32000 | 500 | -.001 | .0248 |

Table 1. Simulations of $\hat{f}$ with $\varepsilon_i = 0$, $F_s = 10MHz$.

From Table.1, one can observe that $Bias\{\hat{f}\}$ can be very small, even if $\Delta f$ is relatively large. One can also see that $Var\{Bias\{\hat{f}\}\}$ becomes smaller when estimating higher frequencies for the same value of $\Delta f$.

## Section 11: Validation of the Mean Square Error $MSE\{\hat{f}\}$

In Section 8 it was shown that even if $\omega$ is only roughly known beforehand (within a few percent), a good $\omega_r$ can be chosen such that $Bias\{\hat{\omega}\}$ can be made negligible. Therefore, assuming $\omega_r$ is in fact chosen reasonably, the additive white Gaussian noise (AWGN) will have much more of a negative effect on $MSE\{\hat{\omega}\}$ and the following is approximately true:

$$MSE\{\hat{\omega}\} = Var\{\hat{\omega}\} + Bias\{\hat{\omega}\}^2 \approx Var\{\hat{\omega}\} \ . \tag{38}$$

Through simulation of the LLS algorithm, this relation will be validated. The simulations had values $\Delta f = -5000Hz$, $q^*$, and AWGN with $SNR = 20dB$.

| $f$ [kHz] | SNRdB | $t_N$ [ms] | Simulation $MSE\{\hat{f}\}$ [$Hz^2$] | Predicted $Var\{\hat{f}\}$ [$Hz^2$] (28) |
|---|---|---|---|---|
| 500 | 20 | 1 | .3107 | .3040 |
| 500 | 20 | .8 | .5937 | .5935 |
| 500 | 20 | .6 | 1.5606 | 1.4072 |
| 500 | 20 | .4 | 4.8247 | 4.7494 |
| 500 | 20 | .2 | 43.2511 | 37.9954 |

Table 2. Result of simulation of $\hat{f}$ with $SNR = 20db$, $F_s = 10MHz$, $\Delta f = -5000Hz$ and using $q^*$.

One can observe in Table.2 that expression (28) is indeed an accurate approximation of the true $MSE\{\hat{f}\}$. Additionally, one can see for $SNRdB = 20dB$, $\hat{f}$ can give a very accurate estimate of $f$ in fractions of a millisecond.

**Section 12: Comparison of Performance with Nonlinear Least Squares (batch)**

In this section, the performance of the batch version of the proposed LLS algorithm is compared with that of a NLS. In these simulations, the same single-tone signal with AWGN was fed to both algorithms.

To calculate the standard deviation of the estimators, the Monte Carlo method [17] was used. 50 signal with AWGN were generated, and the frequency estimates from both algorithms were recorded in vectors. The standard deviation of these vectors was then calculated. These tests were performed for various signal lengths.

| Signal Length $t_N$ [ms] | Standard Deviation of LLS [Hz] | Standard Deviation of NLS [Hz] |
|---|---|---|
| 0.05 | 41.16 | 39.11 |
| 0.1 | 11.13 | 10.71 |
| 0.2 | 3.65 | 3.85 |
| 1 | 0.412 | 0.393 |
| 2 | 0.105 | 0.103 |

Table 3. Results of simulation comparing LLS with NLS, AWGN with $SNR = 27dB$, signal frequency $f = 200kHz$, $\Delta f = 5000Hz$.

From Table 3 it can be seen that the performance of LLS almost exactly matches that of NLS. This is a significant result because LLS is not recursive, requiring vastly less computation resources than NLS. Furthermore, NLS has the tendency to diverge under certain conditions. However, with LLS, this cannot happen, and is thus, a more stable estimator.

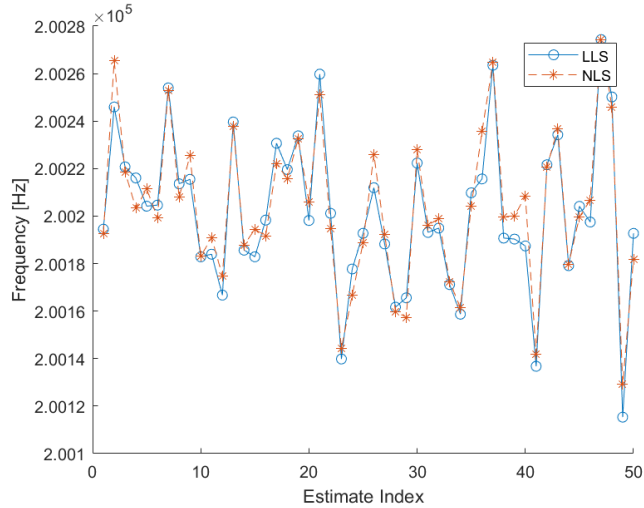

Figure 8. Estimates from a Monte Carlo Test. It can be seen that the results from both estimators are nearly identical.

It can even be observed in Fig.8 that the results of both estimators are nearly identical. This again emphasizes the advantage of LLS for frequency estimation over NLS, since the estimates of LLS were calculated much more efficiently.

**Section 13: Comparison of Performance with a Digital Phase Locked Loop (online)**

In this section, the online performance of the LLS algorithm is compared with that of a DPLL. In these simulations, the settling time of the second order loop filter [11] was chosen to be $2t_N$, which is the maximum time the LLS algorithm would take to fully detect a new frequency shift.

| Settling Time [$ms$] | Standard Deviation of LLS [$Hz$] | Standard Deviation of DPLL [$Hz$] |
|---|---|---|
| 0.05 | 56.62 | 18.33 |
| 0.1 | 22.13 | 6.63 |
| 0.2 | 8.48 | 2.64 |
| 1 | 0.642 | 0.173 |
| 2 | 0.223 | 0.042 |

Table 4. Results of Simulation comparing LLS with DPLL, AWGN with $SNR = 27dB$, signal frequency $f = 400kHz$, . $\Delta f = 5000Hz$

Clearly for the chosen settling time range, the standard deviation of the frequency estimates of DPLL algorithm beats LSS by a factor of about 3. This is still good performance, though any advantage over a DPLL does not lie with the standard deviation. Instead, observe the estimates of both algorithms at the beginning of a simulation.

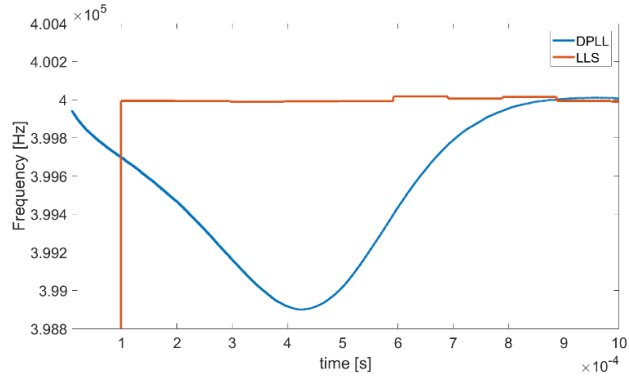

Figure 9. Frequency Estimates of DPLL and LLS, one can observe the Pull-in time of the DPLL.

In the simulation, the Pull-in time of the DPLL was roughly $1ms$. However, one can also see that LLS has no Pull-in time, and begins providing an accurate estimate at time $t_N$.

In fact, LLS has no Pull-in range, or Hold-in range either. This is an advantage over DPLL because it allows for an extremely wide range of frequencies which one can accurately track, while still experiencing similar performance to a DPLL. This behavior can been seen in Fig.9 where the tracked frequency suddenly jumps from $400kHz$ to $405kHz$.

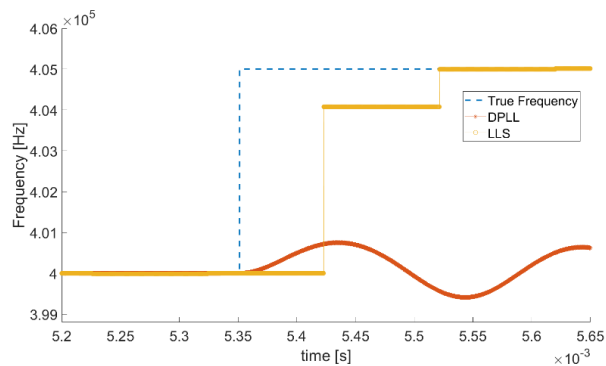

Figure 10. Output of DPLL and LLS when frequency jumps from $400kHz$ to $405kHz$. It can be seen that the DPLL has lost lock.

### Section 14: Conclusion

In this paper an LLS-based algorithm was presented for estimating the frequency of a single tone signal in the presence of white noise. The methodology of this algorithm is to linearize the frequency estimation problem by fitting a waveform of similar, but incorrect frequency. Counterintuitively, by doing this, the correct frequency can then be extracted. Excellent performance was realized when the signal frequency was known beforehand within a range of $\pm 10\%$ or even greater. For example, using the batch version of the algorithm, the standard deviation of the estimator matched that of NLS. This is a significant result because LLS is non recursive, and therefore, much more computationally efficient than the recursive NLS. When LLS was used online, for sub-millisecond settling times this performance was even comparable to that of a Digital Phase Locked Loop, the online frequency-tracking standard in many fields. Furthermore, it was shown that LLS is much more versatile than a DPLL, possessing no Lock-in time , Lock-in range, or Hold-in range.

Appendix 1.

$$E\{\hat{\phi}\} \approx$$

$$\tan^{-1}\left[\frac{-\left[\left(2\omega\cos[\phi]-(\omega+\omega_r)\cos\left[\phi+\frac{2\pi q(\omega-\omega_r)}{\omega_r}\right]+(-\omega+\omega_r)\cos\left[\phi+\frac{2\pi q(\omega+\omega_r)}{\omega_r}\right]\right)(-4\pi q+\sin[4\pi q])+2\sin[2\pi q]^2\left(-2\omega_r\sin[\phi]+(\omega+\omega_r)\sin\left[\phi+\frac{2\pi q(\omega-\omega_r)}{\omega_r}\right]+(-\omega+\omega_r)\sin\left[\phi+\frac{2\pi q(\omega+\omega_r)}{\omega_r}\right]\right)\right]}{\left[\left(-2\left(2\omega\cos[\phi]-(\omega+\omega_r)\cos\left[\phi+\frac{2\pi q(\omega-\omega_r)}{\omega_r}\right]+(-\omega+\omega_r)\cos\left[\phi+\frac{2\pi q(\omega+\omega r)}{\omega r}\right]\right)\sin[2\pi q]^2+(4\pi q+\sin[4\pi q])\left(-2\omega_r\sin[\phi]+(\omega+\omega_r)\sin\left[\phi+\frac{2\pi q(\omega-\omega_r)}{\omega_r}\right]+(-\omega+\omega_r)\sin\left[\phi+\frac{2\pi q(\omega+\omega_r)}{\omega_r}\right]\right)\right)\right]}\right]$$

Appendix 2

$$E\{\hat{\phi}\}\Big|_{q=1/\sqrt{2}} = E\{\hat{\phi}\} = \tan^{-1}\left(\frac{A\cos(\phi)+B\sin(\phi)}{C\cos(\phi)+D\sin(\phi)}\right)$$

$$A = 2\omega\cos\left[\frac{\sqrt{2}\pi\omega}{\omega_r}\right]\left(\sqrt{2}\pi\cos\left[\sqrt{2}\pi\right]-\sin\left[\sqrt{2}\pi\right]\right)+\omega\sin\left[2\sqrt{2}\pi\right]-2\sqrt{2}\pi\left(\omega-\omega_r\sin\left[\sqrt{2}\pi\right]\sin\left[\frac{\sqrt{2}\pi\omega}{\omega_r}\right]\right)$$

$$B = 2\sqrt{2}\pi\omega_r\cos\left[\frac{\sqrt{2}\pi\omega}{\omega r}\right]\sin\left[\sqrt{2}\pi\right]-2\omega_r\sin\left[\sqrt{2}\pi\right]^2+2\omega\sin\left[\frac{\sqrt{2}\pi\omega}{\omega_r}\right]\left(-\sqrt{2}\pi\cos\left[\sqrt{2}\pi\right]+\sin\left[\sqrt{2}\pi\right]\right)$$

$$C = 2\sqrt{2}\pi\omega\cos\left[\frac{\sqrt{2}\pi\omega}{\omega_r}\right]\sin\left[\sqrt{2}\pi\right]+2\omega\sin\left[\sqrt{2}\pi\right]^2-2\omega_r\sin\left[\frac{\sqrt{2}\pi\omega}{\omega_r}\right]\left(\sqrt{2}\pi\cos\left[\sqrt{2}\pi\right]+\sin\left[\sqrt{2}\pi\right]\right)$$

$$D = -2\omega_r\cos\left[\frac{\sqrt{2}\pi\omega}{\omega_r}\right]\left(\sqrt{2}\pi\cos\left[\sqrt{2}\pi\right]+\sin\left[\sqrt{2}\pi\right]\right)+\omega_r\sin\left[2\sqrt{2}\pi\right]+2\sqrt{2}\pi\left(\omega_r-\omega\sin\left[\sqrt{2}\pi\right]\sin\left[\frac{\sqrt{2}\pi\omega}{\omega_r}\right]\right)$$

(39)